# Power Challenges of Large Scale Research Infrastructures: the Square Kilometer Array and Solar Energy Integration

Towards a zero-carbon footprint next generation telescope

Domingos Barbosa
Radioastronomy Group, Inst. Telecomunicações
Campus Universitário Santiago, Aveiro, Portugal
dbarbosa@av.it.pt

Gonzalo Lobo Márquez, Valeriano Ruiz, Manuel Silva
Centro Tecnológico Avanzado Energías Renovables
Paraje los Retamares S/N
04200 Tabernas -Almería, Spain

Lourdes Verdes-Montenegro, Juande Santander-Vela
Instituto de Astrofísica de Andalucia (IAA-CSIC)
Glorieta de la Astronomía s/n, E-18008,
Granada, Spain

Dalmiro Maia, Sonia Antón
CICGE, Dep. de Matemática,
Rua do Campo Alegre nº 687
4169-007 Porto, Portugal

Arnold van Ardenne
ASTRON, P.O. Box 2, 7990 AA Dwingeloo,
The Netherlands
Chalmers University of Technology, Gothenburg
Sweden

Matthias Vetter
Fraunhofer-Institut für Solare Energiesysteme ISE
Heidenhofstraße 2,
79110 Freiburg , Germany

Michael Kramer, Reinhard Keller
Max-Planck für Radioastronomie
Auf dem Huegel 69
53121 Bonn, Germany

Nuno Pereira, Vitor Silva
Logica EM
Parque Tecnológico de Moura, 7860-999 Moura, Portugal

and The BIOSTIRLING Consortium

*Abstract*—**The Square Kilometer Array (SKA) will be the largest Global science project of the next two decades. It will encompass a sensor network dedicated to radioastronomy, covering two continents. It will be constructed in remote areas of South Africa and Australia, spreading over 3000Km, in high solar irradiance latitudes. Solar Power supply is therefore an option to power supply the SKA and contribute to a zero carbon footprint next generation telescope. Here we outline the major characteristics of the SKA and some innovation approaches on thermal solar energy Integration with SKA prototypes.**

## I. Introduction

All future major science infrastructures will have to consider their carbon footprint and Power costs into the respective development path and lifetimes [1]. Additionally, the Roadmap of the European Strategy Forum on Research Infrastructures (ESFRI) [2] has indicated that a multitude of test facilities and Research Infrastructures are paramount to lead the world in the efficient use of energy, in promoting new renewable forms of energy, and in the development of low carbon emission technologies, to be adopted as part of a future Strategic Energy Technology Plan. The Square Kilometer Array (SKA) [3] is an international Information and Computing Technology machine dedicated to radioastronomy that will be built in the Southern Hemisphere in high solar irradiated zones (South Africa with distant stations in Botswana, Ghana, Kenya, Madagascar, Mauritius, Mozambique, Namibia and Australia/New Zealand). It presents an ideal scenario to aggregate renewable energy know-how and become a major Green Infrastructure during its lifetime. In fact, SKA may set a pioneering example for self-sustainable mega-science production and infrastructure operation [3,4,5], with an expected direct economic and indirect societal impacts in the developing nations. The spin-off triggered by SKA will impact on human quality of life opening prospects to provide reliable power to 1.6 billion people and improve the local economy and welfare of remote communities.

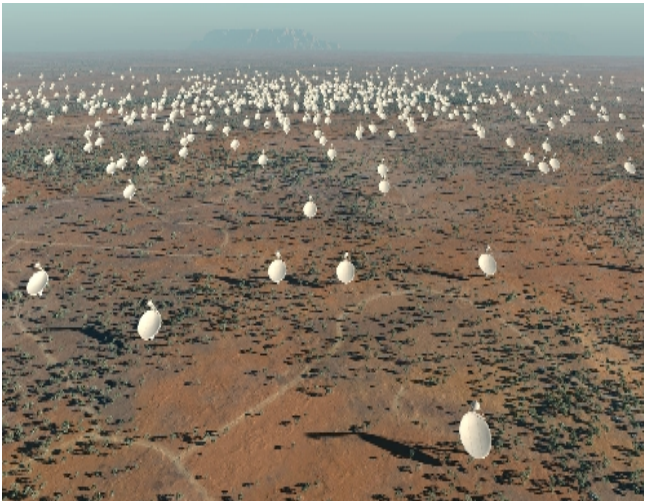

Figure 1. Artist impression of the SKA Core site, with the hundreds of 12-meter parabolic antennas. These antennas shall be connected by power and communication cables. From the core three spiral arms will spawn over 3000km, connecting thousands of antennas and phased array systems (not shown)(http://www.skatelescope.org).

## II. THE SQUARE KILOMETER ARRAY

SKA is a large-scale international science facility involving 67 organizations in 20 countries, as well as world industrial partners and is the only global ESFRI project. The SKA Organization Board includes representatives from Australia, Canada, China, Italy, The Netherlands, New Zealand, South Africa, Sweden, the United Kingdom and soon Germany, officially joining by November 2012. SKA is a multipurpose radio interferometer with thousands of antennas linked together to provide a collecting area of one square kilometer. The SKA can be described a central core of ~200 Km diameter, with 3 spiral arms of cables connecting nodes of antennas spreading over sparse territories up to 3000Km distances. The two SKA sites (South Africa and Australia) were chosen for their exquisitely low Radio Frequency Interference, among other conditions.

Current electronic technology projections point towards an expected target average power usage of the SKA between ~100 MW [5,6,7]. Since SKA will scan the sky continuously, it will not present strong power peaks and power fluctuations, keeping a much smoother consumption profile. Energy generation at a continental scale for this facility, with different load profiles at different locations, means that modular power generators are needed, presenting an ideal scenario for development of innovative solutions with its own degree of customization and grid connectivity. Another consideration is that SKA, by definition, requires 24/7 observation operations. Furthermore, recent experiences with SKA precursors and pathfinders like ASKAP, MeerKAT and LOFAR reveal that an important part of the life cycle cost of these large scale radio astronomy projects will be energy [6,7]. Energy transport infrastructure and generation to supply the electronics and its cooling are key aspects here. This last aspect has a direct impact on system sensitivity to radio signals, as innovative forms of passive cooling plus room temperature operation will have to be considered for the work of the Low Noise Amplifiers of the antenna sensors.

Addressing both the generation of electricity and, on the other side, the reduction of electricity bills is thus paramount to avoid future inefficiencies and higher costs. For instance, ALMA interferometer in the Chilean Atacama desert and the VLT telescope in the Andes use power from diesel generators leading now ESO to consider Green Energy supply to its VLT facilities in the Paranal mountain, in Chile [8]. Here, a quoted example, electricity prices in Chile rose on average by 7% per year between 2003 and 2010 according to statistics from the Organization for Economic Cooperation and Development (OECD) [8,14]. Thus, fossil fuel price fluctuations and longer term availability and associated price rising represent a time challenge in terms of a power grid implementation and consumption. These facts have prompted ASTRON in the context of LOFAR to embark on a solar energy supply strategy investigation.

SKA is planned in two construction Phases, with deployment of different sensor technologies. In particular, performance of digital sensors [1,6] may be driven by the electronics power consumption as power consumption may likely cap sensor performance. Therefore, to extract the maximum scientific potential and maintain costs at appropriate levels it is essential to couple the power cycles with electronic power requirements (power and cooling) in a hot, dry site, with temperatures closer to 50ºC in open field for viable operation. This should be done without degrading receiver system temperature to achieve expected sensitivities, detailed in the SKA Key Science Projects.

Overall, the main characteristic concerning the SKA power system can summarized as :

- Many Antennas nodes are far away from civilization centers and power grid in climates with high thermal amplitudes.

- Exquisite control of Radio Frequency Interference and EMI from Power systems since this would impair the radiotelescope sensitivity.

- Different Power requirements over large distances : the SKA Core ~ 50MW; the High power Computing <50M€; around 80 stations of about 100kW over the spiral arms.

- Continuous operation; ie 24/7 availability meaning some storage capabilities for night operations must be considered.

- Power stability : control of current peaks, for operation, cooling, computing and telescope management and monitoring.

- Scalability: the power infrastructure should scale from SKA Phase 1 to the later Phase2.

TABLE I. POWER PROJECTIONS FOR DIFFERENT SKA SUBSYSTEMS AND PHASES; TARGET : **<100MW.**

| SKA Phase 1 and 2 | South Africa | Australia |
|---|---|---|
| Sparse Arrays | | 3.36 MW |
| Dishes | 1MW | |
| Dishes/PAF | | 0.12 MW |
| On-site Computing | 4.7MW | 1.32MW |
| Totals/site | 5.7MW | 4.8MW |
| SKA Phase2 incl.Dense Arrays | ~80MW (SKA Phase 2 configuration not known yet) | |
| Off-site Computing | ~30-40MW (SKA Phase 2 configuration not known yet) | |

## III. INNOVATIVE APPROACHES: POWERING SKA

### A. Solar Power Sources

The main solar technology supply sources can be divided in (see Figure 2):

- solar thermal energy, collected either by concentrating (towers, parabolic dishes, Fresnel lenses, etc) or flat collectors. Heat can be processed via Stirling engines or via Rankine cycle turbines.

- solar electricity, produced through Photovoltaic panels, plain flat or using concentrating optics (lenses, mirrors); stationary or in trackers

SKA will have a core with 50% of its power needs while the remaining needs will be distributed through the remote stations over 3000Km. Although economic reasoning would prefer a central power plant and a grid infrastructure to power the SKA, yet, to avoid power transport losses over large distances and keeping remote systems self-sufficient, solar power is a preferred option, either through Photovoltaic systems [4,5] or Solar Concentration [16], options covering most solar illumination scenarios (Fig2).

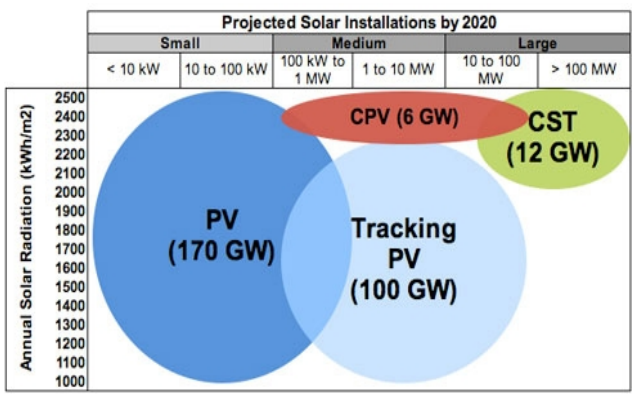

Figure 2. Projections of Solar power market shares by 2020, considering the main technology choices. PV- Photovoltaic; CPV- Concentrated Photovoltaic; CST – Concentrated Solar thermal. From [12].

### B. The BIOSTIRLING 4 SKA

Smaller, but real-scale demonstration projects will provide hints on the best day observing regimes, what best observation strategies can be recommended under this extreme circumstances and how to reconcile peak energy production with peak use and thus peak science production. SKA economic analysis will have to include Power storage and connection to a power supply mix for remote antenna stations and considerations on grid connectivity, autonomous and local storage of power for night operation.

The BIOSTIRLING 4 SKA project [16,17] is led by Spanish companies, involves 14 European partners, with 4 radioastronomy groups to drive specifications. It will study one of the innovative approaches, by deploying parabolic concentrators equipped with Stirling engines. Besides testing an innovative solution to SKA, the BIOSTIRLING project will contribute to the advance of Solar Dish/Stirling technology and hybridization with biomass and energy storage [15], thus optimizing the research and demonstration processes.

The Dish Stirling systems have demonstrated the highest efficiency of any solar power generation system by converting nearly 31.25% of direct normal incident solar radiation into electricity after accounting for parasitic power losses. Therefore, the Dish Stirling technology is anticipated to surpass parabolic troughs by producing power at more economical rates and higher efficiencies. Because the Dish Stirling systems are modular, each system is a self-contained power generator, which can be assembled into plants ranging in size from kilowatts to 10MW. However, the aforementioned technology is not yet commercially exploitable to date as other Concentrated Solar Power (CSP) technologies, such as tower and Parabolic solutions. This is because the current solar dish technology still presents several limitations: high costs, limited life time, low system stability and reliability.

The main targets addressed by BIOSTIRLING in order to validate a new commercial solar dish technology at demonstration scale are:

- **Cost reduction:** developed a new design enabling the mass manufacturing (new manufacturing and O&M Strategies; Structures with less weight with easy commissioning and maintenance).

- **Efficiency:** increase the efficiency by: improving the Stirling engine; develop new reflective materials and new designs with reduction of reflectivity loss.

- **Dispatchability:** The problem of the dependency of an external source like the Sun, will be solved using a combination of two different solutions: hybridization with biomass and energy storage.

- **Life time:** Innovative glass coatings will be developed and new steel with improved resistance and stiffness will be used.

Important for arid locations, there is no need for water supply, except in small amounts for cleaning. The technology demonstrators will be deployed in Moura, in

Southeast Portugal, where SKA related prototypes have been tested. The site has one of the lowest Radio Frequency Interference in Europe, in one the most illuminated geographical area of Europe, by the Southern Iberian border. This optimal characteristics make it ideal to test radioastronomy prototypes with solar demonstrators.

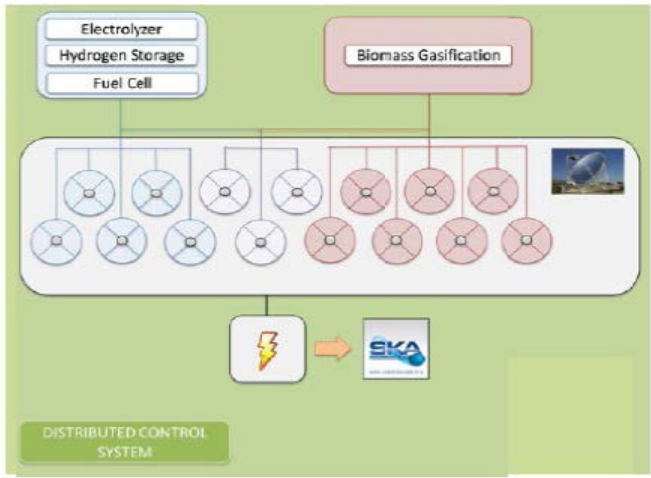

Figure 3. Generic distributed control system of the Stirling concentrators and the hybridization system [16].

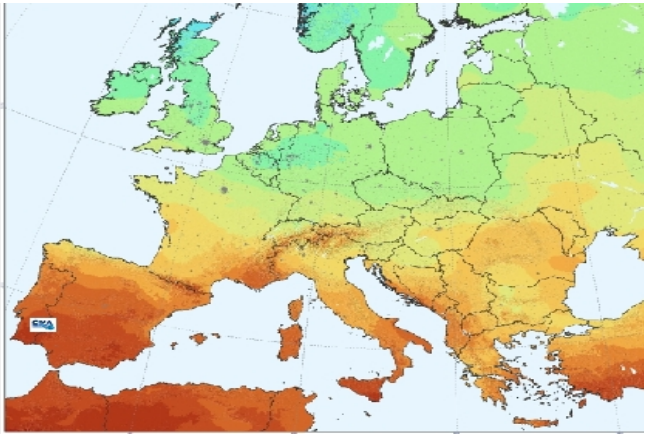

Figure 4. Moura Location in Iberia. Map source: European Committee, IES– Institute for Environment and Sustainability Joint Research Centre.

## IV. CONCLUSION

Solar Energy is a clear power supply option to mega science projects like SKA prompting for the development of innovative power supply mix. Different technologies are available, some of them are very mature others offer promising efficiencies. Among several power options, BIOSTIRLING will study optimization of Dish Stirling engines for a 24/7 operation of radioastronomy and SKA prototype technologies, in Moura, Portugal, closer to the Iberian border region.

ACKNOWLEDGEMENTS

DB, DM and SA acknowledge support from Ciência 2007 and Ciência 2008 Research Contracts funded by FCT/MCTES and POPH/ FSE (EC). VIA-SKA is funded by Ministerio de Economia y Competitividad under Grant AIC-A-2011-0658. JdS and LVM are also funded by Grant AYA2008-06181-C02 and AYA2011-30491-C02-01, co-financed by MICINN and FEDER funds, and the Junta de Andalucía (Spain) grant P08-FQM-4205. We acknowledge Radionet3 and FP7 sponsorship. DB acknowledges sponsorship by TICE and EnergyIn.